\documentclass[twocolumn,showpacs,amsmath,amssymb,prl,graphics,graphicx]
{revtex4}
  \usepackage{graphicx}
\usepackage{bm}
\usepackage{dcolumn}
\begin{document}

\title{Dissipation in ultra-thin current-carrying superconducting
bridges; evidence for quantum tunneling of Pearl vortices}

\author{
F. Tafuri$^1$, J.R. Kirtley$^2$, D. Born$^3$, D. Stornaiuolo$^3$,
P.G. Medaglia$^4$,
P. Orgiani$^4$, G. Balestrino$^4$, V.G. Kogan$^5$}
\affiliation{$^1$INFM-Coherentia Dip. Ingegneria dell'Informazione
Seconda Universit\'{a} di Napoli, Aversa Ê(CE), Italy\\
$^2$IBM Watson research center, Route 134 Yorktown Heights, NY 10598\\
$^3$ INFM-Coherentia, Dipartimento di Scienze Fisiche, Universit{\'a}
di Napoli ``Federico II"
Napoli, Italy Ê\\
$^4$ INFM-Coherentia, Dipartimento di Ingegnaria Meccanica,
Universit{\'a} di Roma Tor Vergata, Roma, Italy \\
$^5$Ames Laboratory - DOE and Department of ÊPhysics and Astronomy,
Iowa State University, Ames IA Ê50011-3160
}

\date{\today}
\begin{abstract}
We have made current-voltage (IV) measurements of
artificially layered high-$T_c$ thin-film Êbridges.
Scanning SQUID microscopy of these films
provides values for the Pearl lengths $\Lambda$ that exceed
the bridge width, and shows that the current distributions
are uniform across the bridges. At high temperatures and high currents the
voltages follow the power law   $V \propto I^n$,
with $n=\Phi_0^2/8\pi^2\Lambda k_B T+1$, and at high temperatures and
low-currents
the Êresistance is exponential in temperature, in good
agreement with the predictions for thermally activated vortex motion.
At low temperatures, the IV's
are better fit by $\ln V$ linear in $I^{-2}$.
This is expected if the low temperature dissipation is dominated by
quantum tunneling of Pearl vortices.
\end{abstract}
\maketitle

Soon after the seminal work of Caldeira and Leggett \cite{Leggett}, the
possibility of quantum tunneling of vortices in thin films was
considered by Glazman and Fogel \cite{Glazman}. This idea was
studied extensively in theory and experiment in relation Êto
{\it finite} flux-creep rates Êin hard type-II superconductors for
$T\to 0$. For thermally activated
creep, the rates should go to zero in this limit, see review
\onlinecite{Blatter} and references therein. Within Êcreep models, the
barriers through which the vortices tunnel are due to material disorder and
are treated statistically.

Narrow current carrying Êthin-film bridges offer a
unique opportunity to study both
thermal activation and quantum tunneling through Êthe
  well defined potential barriers that are tunable
by current and temperature.
Within the London approach, Êthe barrier shape is well known for
bridges narrow relative to the Pearl length
$\Lambda=2\lambda^2/d$, where $\lambda$ is the penetration depth and $d$ is
the film thickness.

To realize these conditions we use ultra-thin high-$T_c$ films with extremely
large Pearl lengths $\Lambda\sim 100\,\mu$m. The critical temperature of these
films $T_c\sim 40\,$K provides a relatively easy $T$-domain to work in, and the
data reported are quite robust.

When current $I$ flows through the bridge at high temperatures,
vortices are Êthermally activated at the strip edges and pushed in,
causing dissipation and a non-zero voltage $V$. We show that these
processes should lead to a power-law dependence
$V\propto I^n$ Êwith the exponent $n$ determined by the film parameter
$\Lambda$ and temperature:
\begin{equation}
n=\frac{\phi_0^2}{ 8\pi^2\Lambda(T) k T}+1\,,\label{n}
\end{equation}
($\phi_0$ is the flux quantum and $k $ is the Boltzmann constant).
This Êis Êconfirmed by the high-temperature  IV's and by the
independently measured Pearl length.

However Êfor low $T$'s, the
IV's show a slower current dependence than required by
thermal activation. Instead, a quantum tunneling model utilizing
the ideas of Refs.\,\onlinecite{Leggett,Glazman},
along with our knowledge of the barrier shape, provide a good
representation of the data.

In the present work, ultrathin sandwiches of
[Ba$_{0.9}$Nd$_{0.1}$CuO$_{2+x}$]$_{5}$/[CaCuO$_{2}$]$_{2}$/[Ba$_{0.9}$Nd$_{0.1}$CuO$_{2+x}$]$_{5}$
(CBCO 5/2/5) were used. They consist of one superconducting infinite
layer block (two CaCuO$_{2}$ unit cells), sandwiched between two charge
reservoirs ($5$\ Ba-based unit cells). The structure was grown on (001)\
SrTiO$_{3}$ substrates, with nominally zero miscut angle, by
  using a focussed KrF excimer pulsed
laser source \cite{balestrino0,balestrino apl}. The films were
photolithographically patterned into bridges. In this paper we present
measurements from a ``wide" bridge approximately 85$\mu$m across. After
measurement, this bridge was   narrowed by a factor
of 2 and then remeasured. The patterning of such a thin layer
(5\,nm with a superconducting cell  only 1\,nm thick) protected by a 100\,nm
amorphous cap layer required specially careful
  photolithography and Êmilling. The
use of a cap layer deposited in situ required pre-deposited Au
contacts (details will be given elsewhere). Four terminal
measurements of the current-voltage characteristics
were realized down to 300\,mK in an Oxford Heliox probe suitably
modified by superconducting wiring to minimize heating at elevated
currents. The bridges were
 imaged with a scanning SQUID microscope (SSM)
\cite{ssmapl,vartapl} to determine
current uniformity and the film Pearl length.

\begin{figure}[htb]
\includegraphics[ width=3.5in]{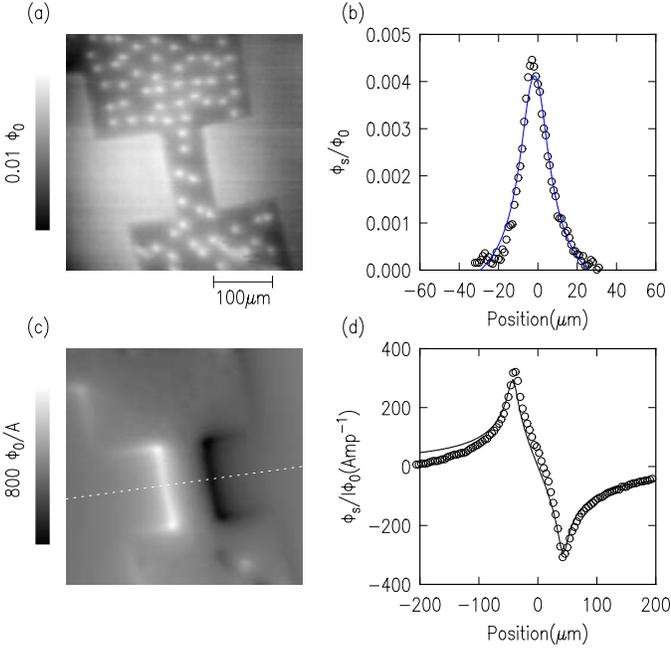}
\caption{\label{fig:brdgecrs}(Color online) 
(a) SSM image of the wide 5/2/5 CBCO bridge
after cooling in a 22\,mG field Êat 5\,K with an 8$\,\mu$m square
pickup loop. (b) Cross-section through an isolated Pearl
vortex after cooling in a 1\,mG field. The line fit is obtained using
the field distribution $h_z(x,z)$ given in Ref.\,\onlinecite{cbcoprl},
with an effective pickup loop height $z=5\,\mu$m and
$\Lambda =400\,\mu$m. (c) SSM image with a 104\,Hz, 100$\,\mu$A r.m.s.
a.c. current passing through the bridge at $T=$20\,K.
(d) Cross-section through the data along
the dashed line in (c); the solid line is calculated for uniform current
through the bridge, with $z=5 \,\mu$m.}
\end{figure}

Figure \ref{fig:brdgecrs}a shows an SSM image of the wide bridge
after cooling in a 22\,mG field. The bridge outline is visible along
with Êtrapped Pearl vortices \cite{pearlapl,cbcoprl}. Fig.
\ref{fig:brdgecrs}b shows a cross-section through one Pearl vortex,
after recooling in a smaller field to reduce the density of vortices.
Fitting such cross-sections results in low temperature Pearl lengths of
400$\pm$50$\,\mu$m for this bridge. Fits to scanning susceptibility
measurements   of the same bridge gave $\Lambda=$
200$\pm$20$\,\mu$m (see Ref.\,\onlinecite{cbcoprl} for details of these
procedures). This difference may be due to the finite lateral dimensions
of the film, which are not accounted for in the modelling.
The length $\Lambda$ is therefore longer than the bridge
width, and we expect the current through the bridge to be uniform, as
confirmed by the SSM image and fit of Fig.\,\ref{fig:brdgecrs}c,d. The
field $h_z$ (normal to the film) is maximum near the edges $x=0,W$ where
vortices Êor antivortices Ênucleate, penetrate the strip by being pushed in
by the Lorentz Êforce $\phi_0
I/cW$, and Êannihilate near the strip middle. We focus on the barrier for
vortices; the one for antivortices is obtained by replacing $x$ with $W-x$.

For narrow strips of width $W\ll \Lambda$ with no current, the
energy of a vortex at a position $0<x<W $ is \cite{Pearlvortex}:
\begin{equation}
\epsilon(x) =
\frac{\phi_0^2}{8\pi^2\Lambda}\ln\left(\frac{2W}{\pi \xi}\sin
\frac{\pi x}{W}\right)\,,
\label{energy}
\end{equation}
$\xi$ is the coherence length. 
The vortex energy in the presence of a uniform current is
\begin{equation}
U(x)=\epsilon(x) -\frac{\phi_0 I}{cW}\,x \,.
\label{potential}
\end{equation}
The current causes suppression of the barrier maximum $U_m$ and pushes
the maximum position $x_m$ from the middle (at $I=0$) toward the edge
$x=0$. We obtain Êafter simple algebra:
$x_m= (W/\pi)\,\tan^{-1}(I_0/I)$ and
\begin{eqnarray}
U_m=\epsilon_0\left( \ln \frac{2W}{\pi
\xi\sqrt{1+ I^2/I_0^2}} Ê-
\frac{I}{I_0}\tan^{-1}\frac{I_0}{I}\right),
\label{Um}\\
I_0=\frac{c\phi_0}{8\pi\Lambda}\,,\qquad
\epsilon_0=\frac{\phi_0^2}{8\pi^2\Lambda}\,. \label{Io}
\end{eqnarray}
For $\Lambda\approx 400\,\mu$m, $I_0\approx 2 \times 10^{-3}\,$mA at low
temperatures and decreases on warming. For the majority of our data
$I/I_0\gg 1$ and Eq.\,(\ref{Um}) simplifies to:
\begin{eqnarray}
U_m=\epsilon_0\ln\frac{2 WI_0}{e\pi\xi \, I
 }   \,,\label{barrier}
\end{eqnarray}
where $e=2.718$. The logarithmic dependence of the barrier height
on the current results in power-law  IV's, the dependence
encountered in models of vortices moving through
random potentials in disordered materials \cite{Blatter}. In our
case, it is caused by a well-defined barrier.


If vortices cross the strip via a thermally
activated process, a voltage $V\propto \exp(-U_m/kT)$ is generated,
i.e.,
\begin{eqnarray}
V &\propto& \exp\left(-\frac{\epsilon_0}{kT} \ln \frac{
2WI_0}{e\pi\xi ÊI
 } \right)   = \left(\frac{I}{I_d}\right)^{m}
\,,\label{V}\\
m&=&\frac{\epsilon_0(T)}{kT}=
\frac{\phi_0^2}{8\pi^2\Lambda(T)kT }\,,\\
I_d&=&\frac{c\phi_0}{8e\pi^2\lambda^2\xi}\, Wd\,.\label{Idp}
\label{n_estimate}
\end{eqnarray}
where $I_d$ is of the order of the bridge depairing current.
Thus, the model provides not only the power-law {\it per se}, but gives the
exponent in terms of the film parameters. We note also that our
model differs from that used in Ref.\,\onlinecite{Repaci} to 
interpret similar data.
 
\begin{figure}[htb]
\includegraphics[ width=70mm]{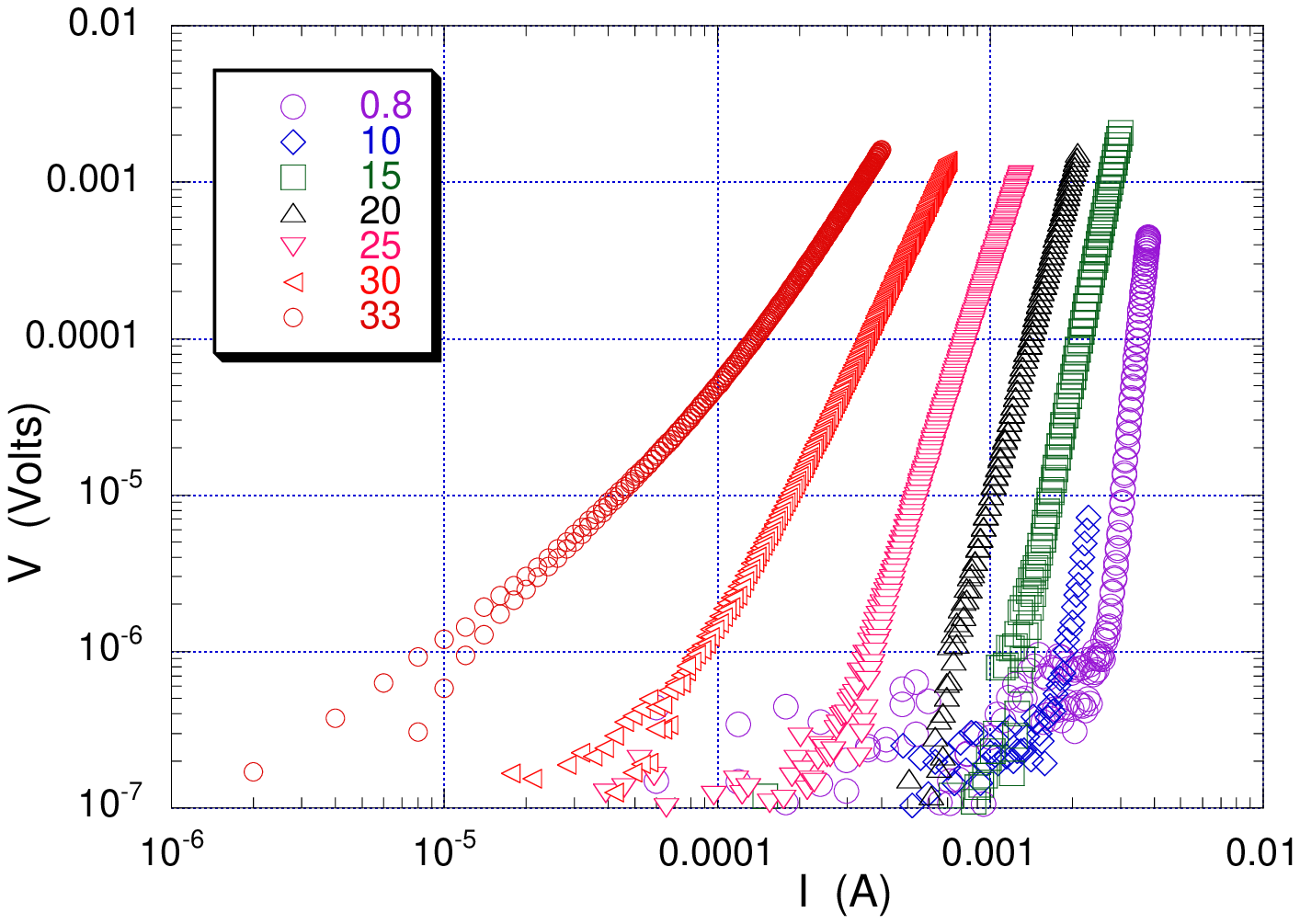}
\includegraphics[ width=65mm]{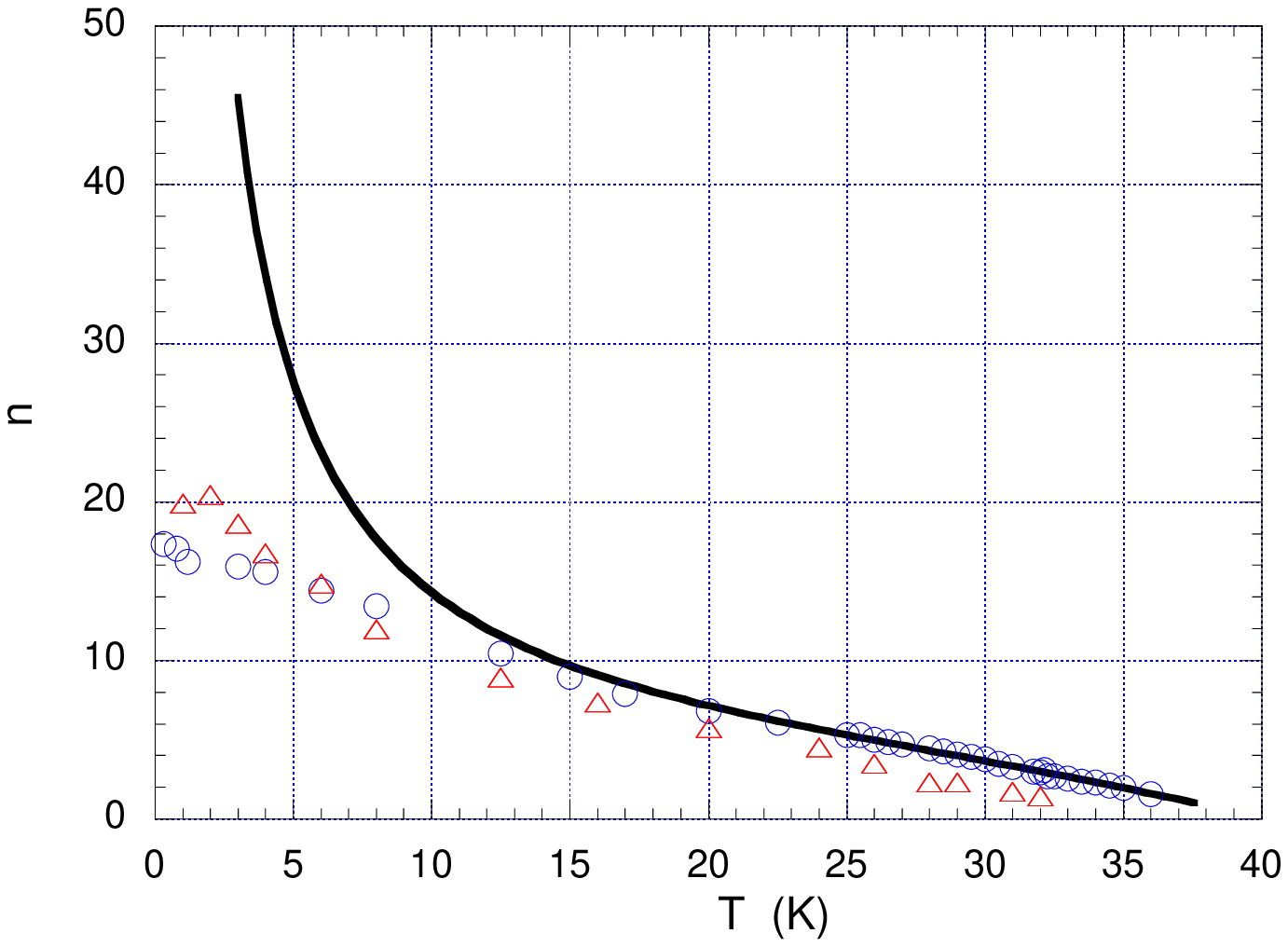}
\caption{\label{f1}(Color online) The upper panel: $V(I)$ on a log-log
scale at temperatures indicated in the legend. ÊLower panel: The exponent
$n$ extracted from $V\propto I^n$ for the wide bridge (circles) and the
narrow bridge (triangles). The solid line is $n(T)$ of Eq.\,(\ref{n})
with with the best-fit parameters $T_c=37.6\,$K and
$\Lambda(0)=317\,\mu$m. Ê}
\end{figure}

As is usually the case with thermal activation, it is difficult to fix
the ``attempt frequency" Êin the probability $\exp(-U_m/kT)$. Still, the number
of vortices penetrating the strip must be proportional to the field value at
the edges, i.e., to the current
$I$. Adding a prefactor const$\cdot I$ in Eq.\,(\ref{V}), we
obtain the exponent $n=m+1$ of Eq.\,(\ref{n})
which gives the correct limit for $T\to T_c$.

The upper panel of Fig.\,\ref{f1} shows IV measurements for the wide bridge
on a log-log scale. All curves are
nearly straight (except noisy low-voltage parts), i.e., $V\propto I^n$
at high currents. The exponent $n$ extracted from power-law fits to the data is
shown in the lower panel, for both the wide and narrow bridges. The solid line
is $n(T)$ of Eq.\,(\ref{n}), in which the two-fluid
$\Lambda=2\lambda_0^2/d(1-t^4)$ has been used for simplicity;
$t=T/T_c$.
 Fitting the high-$T$ part of the data for the wide bridge we obtain
$T_c\approx 38\,$K and $\Lambda(0)\approx 320\,\mu$m in   agreement with
$\Lambda(0)\approx 200-400\,\mu$m independently measured by SSM. Thus,   the
thermal activation model works well   above
$\approx 15\,$K, where it gives Êcorrect values of the exponent $n(T)$.
Although $T_c$ Êof the narrow bridge was slightly lower after repatterning, the
high temperature exponents were quite close, which means that the Pearl
length was nearly the same in both bridges.

\begin{figure}[htb]
\includegraphics[ width=65mm]{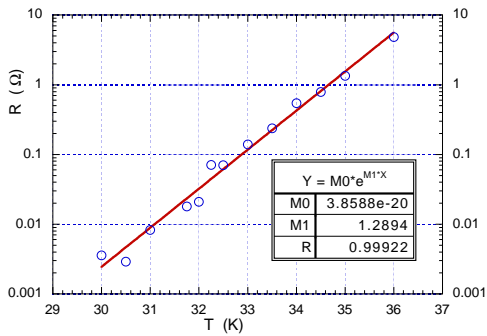} 
\caption{\label{f5}(Color online) $R(T)$ for $30<T<36\,$K and fit
to an exponential dependence.   }
\end{figure}

Another quantity which can be extracted from the IV's is the
low-voltage differential resistance, defined as $R=(dV/dI)_{I\to 0}$.
At low temperatures, the
voltages at low currents are exceedingly small and cannot be extracted
from the noise. However, reliable determination of
$R$ is possible above roughly 30\,K. Fig.\,\ref{f5} shows that in this
domain Ê$R\propto \exp(T/T_0)$.

To understand this result, we turn again Êto the voltage expression
(\ref{V}) for thermally activated vortex motion. As mentioned above,
the prefactor in this expression is
proportional to the current Êbecause the number of vortices attempting
to overcome the barrier is proportional to the edge fields,
i.e., to $I$. Therefore, calculating
$R=(dV/dI)_{I\to 0}$, we can leave the current dependence only in the
prefactor and set $I=0$ in Ê$\exp[-U_m(I)/kT]$. We then have:
\begin{equation}
U_m\approx \epsilon_0
\ln\frac{2W}{\pi\xi}=\frac{4\phi_0^2(1-t)}{8\pi^2\Lambda(0)}
\ln\frac{2W}{\pi\xi}\,,
\end{equation}
where $\Lambda(0)$ is the low temperature Pearl length,
  and the two-fluid approximation
$\Lambda =\Lambda(0) /(1-t^4)\approx \Lambda(0) /4(1-t)$ is again used. Then,
the voltage and the resistance near $T_c$ should be proportional to
\begin{eqnarray}
\exp\left(+\, \frac{
\phi_0^2\,T}{2\pi^2\Lambda(0)k T_c^2}\ln\frac{2W}{\pi\xi}\right).
\end{eqnarray}
Thus, we expect the behavior $R\propto \exp(T/T_0)$ with
\begin{equation}
T_0=\frac{2\pi^2\Lambda(0)k T_c^2}{\phi_0^2\ln(2W/\pi\xi) Ê}
\,, \label{To}
\end{equation}
where $T_0$ Êdepends weakly on $T$ via $\ln \xi$. ÊNote that our
model does not allow us to literally take the limit $T\to T_c$ because
the   energy expression (\ref{energy}) is valid only for $W\gg\xi$.
The data of Fig.\,\ref{f5} correspond Êto $T_0\approx 0.78\,$K.
Expression (\ref{To}) yields a smaller value $T_0\approx 0.3\,$K.

We stress that Ê$R\propto \exp(T/T_0)$ is obtained here
for the thermally activated process, so that this Êdependence {\it per
se} is not necessarily a Êmanifestation of quantum tunneling (compare
with \cite{Goldman}). 

At low temperatures, the thermal activation model fails.
The fit values for $n(T)$ in Fig. \ref{f1}
do not diverge as thermal activation requires Ê\cite{remark1}.
Hence, we turn to the possibility of quantum tunneling. ÊAccording to Ref.
\onlinecite{Glazman}, the tunneling probability for {\it overdamped}
processes   is proportional to
$\exp\left(- \gamma\eta\, x_b^2/\hbar \right)$ Êwhere
$\gamma$ is a geometric factor related to the barrier shape,
$\eta$
is the drag coefficient,
and $x_b$ is the barrier {\it
width}. Unlike thermal activation, for which only the barrier height is
relevant, for the tunneling Êthe barrier width dominates the probability.

Since the potential $U=0$ Ê\cite{remark2} at the position of vortex
entry, the width $x_b$ is a root of $U(x)=0$, or of
\begin{equation}
\ln\left(\frac{2W}{\pi \xi}\sin
\frac{\pi x}{W}\right) = \frac{I}{I_0}\,\frac{\pi x}{W}\,.
\label{e10}
\end{equation}
As argued above, $I/I_0\gg 1$; therefore Ê$\pi x/W$ must be small.
 Since the log is a   slow function, we have
\begin{equation}
x_b\approx C_0\,\frac{WI_0}{\pi I}\,,
\label{C/i}
\end{equation}
where $C_0$ is a constant of the order one.
Hence, the barrier width
shrinks with increasing current. One can show by solving numerically
Eq.\,(\ref{e10}) that for   $I/I_0\sim 10^3$ (the range of our data)
$C_0\approx 5$.

We then expect the voltage for high currents to behave   according to
\begin{equation}
\ln V =C_1-C_2\,\frac{\eta W^2}{\hbar}\,\frac{I_0^2}{I^2}\,.
\label{quant}
\end{equation}
Here, $C_1$ is a constant related to the Êattempt frequency, whereas
$C_2=C_0^2\gamma/\pi^2 \sim 2.5\gamma$.

\begin{figure}[htb]
 \includegraphics[ width=70mm]{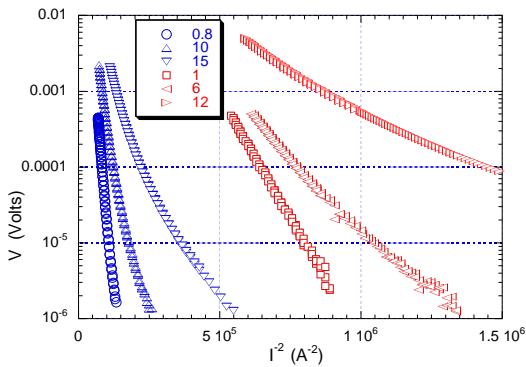}
 \caption{\label{f4} (Color online) The voltage $V$ {\it versus} $1/I^2$ for
 temperatures indicated in the legend. The left group of IV's is for
the wide bridge with $W=85\,\mu$m; the right group is for
$W=42.5\,\mu$m.}
 \end{figure}

In Fig.\,\ref{f4}, we plot log$V$ for a few   $T$'s {\it versus}
$I^{-2}$ for both wide and narrow bridges. Clearly, the straight parts
of the IV's for low Êtemperatures and high currents Êare in agreement with
Eq.\,(\ref{quant}). With increasing $T $, the IV's curve Êaway
from a straight line much faster, signalling deviations from
predominantly quantum tunneling.
We fit the data for $0.8\,$K of the wide bridge and for $1\,$K of
the narrow one to
$V=V_0\exp(-\alpha/I^2)$ to obtain
$V_0$=0.43\,V,
$\alpha=1.0\times Ê10^{-4}$A$^{2}$ for the first and
$V_0$=0.97\,V,
$\alpha=0.14\times 10^{-4}$A$^{2}$ for the second.

Going back to Eq.\,(\ref{quant}) for the quantum tunneling, we note
that the coefficient of $I^{-2}$ is proportional to $W^2$. Therefore,
as an extra check of applicability of this equation, one can Êcompare
$\alpha$'s of the two bridges. Given the approximate character of
our model, the {\it reduction} of $\alpha$ by a factor of 7 in the
bridge $W/2$ as compared with the   expected   {\it reduction}
by a factor of 4, shows that quantum tunneling is consistent with our
data.

Moreover, the pre-exponential factor $V_0$ proportional to the
attempt frequency should go as $1/W$ because - as
magnetostatics shows - the edge field is $\propto 1/W$. The ratio of
$V_0$'s for the two bridges is 0.97/0.43=2.3 instead of the expected
factor of 2. This is yet another indication that the low-$T$ data
are consistent with the tunneling model.

Further quantitative tests of the quantum tunneling model would
be Êpossible if the drag coefficient $\eta$ and the barrier shape
factor $\gamma$ were well known. Unfortunately, this is not the case.
The factor $\gamma=\pi/2$ was found for a barrier $U(x)$ as a
  cubic parabola \cite{LO}, quite different from our case.
For $I_0=2.5\mu A$ and
W=85$\mu$m, Êwe estimate $\eta\gamma\sim 10^{-16}\, $CGS. If one takes
$\gamma\sim 1$, this gives the drag coefficient
 about two orders of magnitude smaller than the Bardeen-Stephen
value $\eta=\phi_0^2d/2\pi\xi^2c^2\rho_n$. Kramer-Pesh
contraction \cite{KP} of the vortex core at low $T$'s might be a
cause of the $\eta$ suppression.

  During this work, we have also searched for a
Berezunskii-Kosterlitz-Thouless transition in the domain of the
power-law IV's. We, however, did not detect any discontinuities in the
$T$ dependence of $n$. We believe this is because in
bridges Êwith $W<\Lambda$,   interaction of vortices is
short-range with a cutoff at distances $W$.


To summarize, at high temperatures we observe the power-law IV's   well
described by thermally activated motion of vortices through the
thin-film strips narrow on the scale of Pearl
$\Lambda$ for which the shape of the potential barrier is well established.
The predicted dependence of the exponent $n$ on
$\Lambda(T)$ is in good agreement with the data. The Êlow temperature
  data are consistent with the overdamped quantum tunneling
of Pearl vortices, the size of which in our films
is Êindeed macroscopic since $\Lambda\sim   Ê0.1\,$mm.

\begin{acknowledgments}
The authors thank L. Bulaevskii for thoughtful remarks. 
This work has been   supported by   ESF projects 
``Pi-Shift"  and ``QUACS", and by MIUR program ``Quantum
effects in Nano-structures and Superconducting devices".

\end{acknowledgments}


\begin{thebibliography}{99}
\bibitem{remark1} Taking a faster than the
two-fluid low-$T$ behavior of $\Lambda$ (for materials with gap
nodes) would have made the increase of $n(T)$ at low $T$'s even stronger.
The levelling of $n(T\to 0)$ is not Êpronounced in
other reports, see e.g. Ref.\,\onlinecite{Repaci}.
\bibitem{remark2} One can take the vortex energy   Êat $x\sim\xi$
near the edge, $\epsilon(\xi)=\epsilon_0\ln 2$, as the starting
energy. This results only in a renormalization of the large constant
$2W/\pi\xi \to W/\pi\xi$, but leaves the structure of Eq.\,(\ref{e10})
unchanged.
\end{thebibliography}
\end{document}